\documentclass[aps,pre,reprint,superscriptaddress,showpacs,showkeys]{revtex4-1} 
\usepackage[utf8]{inputenc}

\usepackage{cmap} 
\usepackage[T1]{fontenc}
\usepackage{microtype}
\usepackage{amsmath,amssymb,mathtools}
\usepackage{txfonts}

\usepackage{mathrsfs}
\usepackage{mleftright}

\usepackage{graphicx,grffile}

\usepackage{textcomp}
\usepackage{siunitx}
\usepackage{booktabs,tabularx,dcolumn}
\newcolumntype{d}[1]{D{.}{.}{#1}}

\usepackage{url,hyperref}
\usepackage[usenames,dvipsnames]{xcolor}
\hypersetup{colorlinks=true, linkcolor=BrickRed, urlcolor=blue!50!black, citecolor=blue!50!black}

\usepackage[capitalize]{cleveref}
\crefrangelabelformat{equation}{(#3#1#4)$-$(#5#2#6)}

\renewcommand\rho\varrho
\renewcommand\vec[1]{\textrm{\bfseries #1}}

\begin{document}
\title{Coalescence preference and droplet size inequality during fluid phase segregation}

\author{Sutapa Roy}
\affiliation{Max-Planck-Institut f\"{u}r Intelligente Systeme,
   Heisenbergstr.\ 3,
   70569 Stuttgart,
   Germany,
   and \\
   IV. Institut f\"{u}r Theoretische Physik,
   Universität Stuttgart,
   Pfaffenwaldring 57,
   70569 Stuttgart,
   Germany}

\date{\today}

\begin{abstract}
Using molecular dynamics simulations and scaling arguments, we investigate the coalescence preference 
dynamics of liquid droplets in a phase-segregating off-critical, single-component fluid. 
It is observed that the preferential distance of the product drop from its larger parent, during a 
coalescence event, gets smaller for large parent size inequality. 
The relative coalescence position exhibits a power-law dependence on the parent size ratio with an 
exponent $q \simeq 3.1$. This value of $q$ is in strong contrast with earlier reports $2.02$ and $5.01$ 
in the literature. The dissimilarity is explained by considering the underlying coalescence mechanisms.

\end{abstract}
\maketitle

\section{Introduction}
When two liquid droplets come in contact with each other they form a liquid bridge and the composite 
structure finally relaxes to a single big drop -- a kinetic process known as coalescence. In recent 
years this phenomenon has gained significant research attention for a wide variety of natural systems 
including collision and coalescence of droplets on a solid surface \cite{gaskell2007, wasserfall2017, narhe2008, zhua2017,
koplik2015}, 
coalescence of water drops in water \cite{malfreyt2013}, collision of 
rain drops \cite{wilkinsin2016}, magneto coalescence of ferrofluidic drops \cite{kadivar2014}, etc. 
Most of the studies on droplet coalescence sought to understand the growth of liquid bridge, 
and the effects of contact angle. On the other hand, 
an intriguing yet rarely explored feature of coalescence processes is the so-called 
\textit{coalescence preference}: the product drop (bubble) which emerges from the coalescence 
of two different sized parent droplets (bubbles) tends to be placed closer to its larger parent. 
Such a preferential positioning is caused by the Laplace pressure 
\cite{tadrosbook} difference between the parents. A smaller parent has higher Laplace pressure compared 
to the larger one and hence the merged product is formed closer to its larger parent. 
While this general trait is conceivable, understanding of the spatial and temporal properties of 
the coalescence preference effect is still in infancy and 
very recent \cite{weon2015, je2012}. Specifically, the questions of `how close' to the larger parent 
the merged drop forms and its dependence on the parent size ratio are unsettled issues \cite{weon2015} and calls for 
future studies. The microscopic mechanism of coalescence preference also remains 
poorly understood.   

In this paper, using extensive molecular dynamics (MD) simulations we investigate 
the droplet coalescence preference in phase-segregating fluids which are rendered 
thermodynamically unstable via a sudden change of temperature.
In recent years, droplet coalescence and growth in phase-segregating \cite{puri2009, bray2002} fluids has gained 
huge momentum because of the underlying rich physics originating from the combined effects of 
hydrodynamics and diffusion field \cite{roy2012, shimizu2015, oprisan2017}, and their temperature and 
dimensionality dependence \cite{das2017}. However, the coalescence preference (CP) dynamics 
in such systems has never been investigated at all. 
On the other hand, fluids coarsening \cite{binder2009, onuki2002} via droplet / bubble coalescence 
are very common in nature. To the best of our knowledge, this is the first study of coalescence 
preference effects in a system close to phase transition.
Apart from having fundamental importance such studies also hold particular 
relevance in industrial applications. For example, the stability of emulsions / foams which is a crucial 
factor for pharmaceutical and petroleum industries is controlled by 
droplet / bubble coalescence \cite{pagureva2016}. For an improved prediction and control of emulsion 
stability it is necessary to understand the geometric preference and the microscopic mechanism 
of re-arrangement during coalescence \cite{biance2011}.

\begin{figure}
\includegraphics*[width=0.42\textwidth]{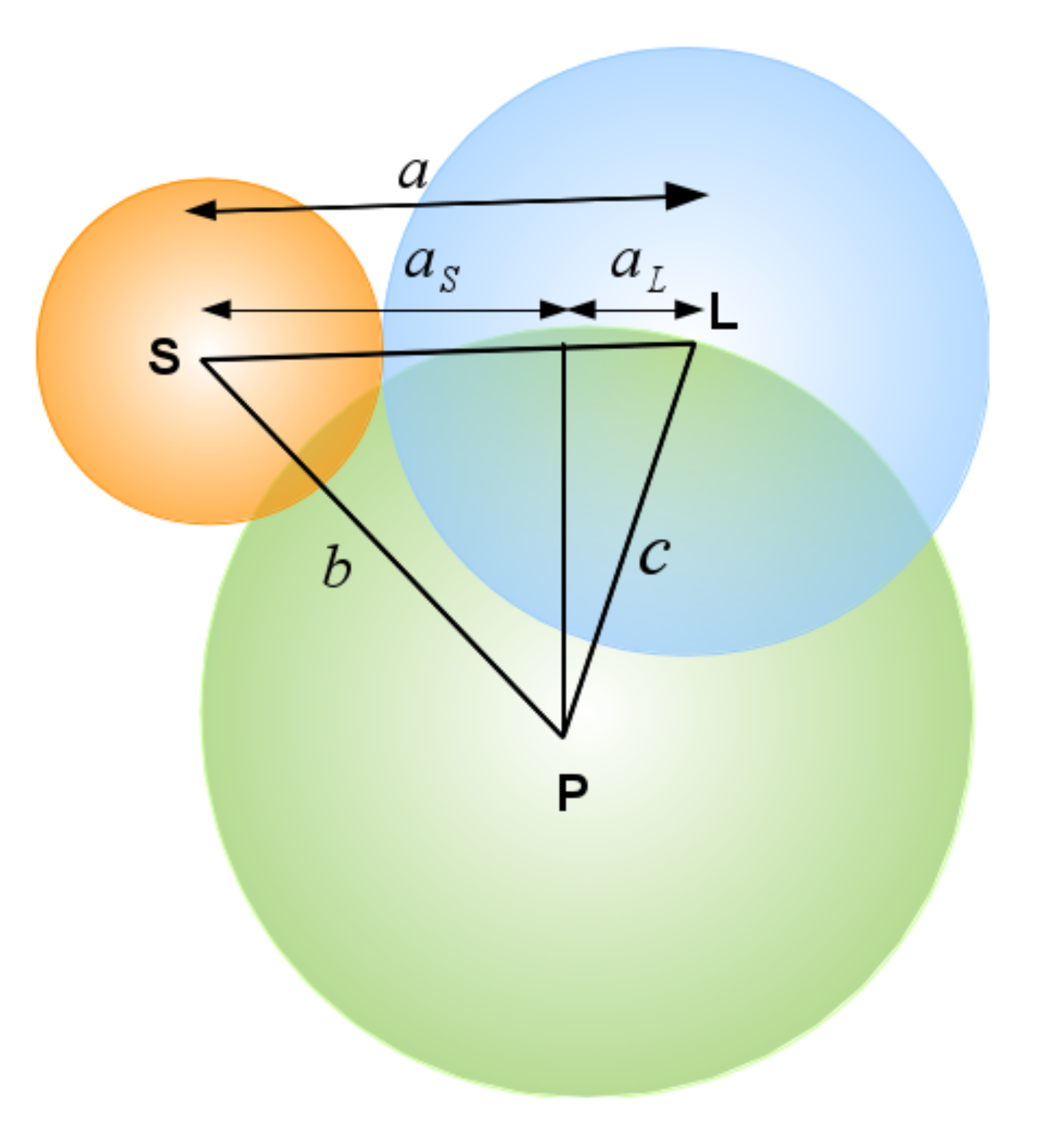}
\caption{\label{fig1}Schematic of two different size coalescing droplets with centres S and L and 
the product drop with centre P. Pairwise distances between the parent and product droplets 
are denoted by $a$, $b$, and $c$. $a_L$ and $a_S$ are separation distances between larger and smaller 
parent, respectively, from the point on the line linking S $\&$ L closest to P.
}
\end{figure}

Typically, proximity of the merged drop to its larger parent is dictated by the 
relative coalescence position $a_L/a_S$ (see \cref{fig1}); where $a_L$ and $a_S$ are the 
distances of the centres of the larger and smaller parent, respectively, to the projected centre of 
the coalesced drop on the line joining the two parents. As evident from recent experiments 
\cite{weon2015, je2012}, the location of the product drop depends on the parent size ratio 
$R_L/R_S$ in a power-law fashion
\begin{equation}\label{ratio2}
 \Big(\frac{a_L}{a_S}\Big) \sim \Big(\frac{R_L}{R_S}\Big)^{-q},
\end{equation}
$R_L$ and $R_S$ being the radii of the larger and the smaller parent, respectively. 
The CP exponent $q$ determines the closeness of the merged drop to the 
larger parent. For a fixed parent size ratio, a higher $q$ means the product forms more 
closer to the larger parent.

As will be demonstrated, our results on phase-segregating fluids reveal that 
with increasing size ratio of two coalescing droplets the product drop 
forms much closer to the larger parent. Although this general trend is similar to earlier 
reports \cite{je2012, weon2015} on droplets / bubbles, the ``extent'' of 
closeness of the merged drop to the bigger parent and the underlying 
mechanism of coalescence preference, as observed in our study, are strikingly different from 
previous reports. In particular, the observed exponent $q \simeq 3.1$ is in 
strong contrast with earlier findings $q \simeq 2.1$ and $5.1$. 
This difference has been attributed to the underlying coalescence 
mechanism. For phase segregating fluids where the coalescing drops undergo 
inelastic sticky collisions \cite{binder1974} the CP dynamics is governed by the formation of the product drop 
at the \textit{centre of mass} location of the parents. This leads to a value 
$q \simeq 3.0$. Whereas, higher values of $q$ in \cite{je2012} were explained to be 
due to the coalescence motion controlled by the release of surface energy.

We first explain the CP dynamics here. \Cref{fig1} illustrates schematically one coalescence event during fluid 
phase separation, where the 
centres of the smaller, larger parent and the product drop are marked by S, L, P, respectively. 
In order to describe the relative position of the product drop we define $a_L (a_S)$ as the distance between 
the larger (smaller) parent centre and the point on the line linking the two parent centres closest to the 
product centre. Using geometric considerations \cite{weon2015} the relative position $a_L/a_S$ can be directly measured from 
the locations of the drop centres as 
\begin{equation}\label{ratio1}
\Big(\frac{a_L}{a_S}\Big)=\frac{\big(a^2-b^2+c^2\big)}{\big(a^2+b^2-c^2\big)},
\end{equation}
where, $a$, $b$, $c$ are defined in \cref{fig1}. \Cref{ratio1} involves only the pair-wise distance among 
the three droplets and therefore, knowledge on the droplet positions suffices the calculation of ${a_L}/{a_S}$. 
From earlier experiments \cite{weon2015, je2012}, ${a_L}/{a_S}$ is known to exhibit an algebraic dependence on the parent 
size ratio as dictated by \cref{ratio2}. Note that for equal size droplets the product always forms exactly at the centre 
of the two parents, i.e., $a_L/a_S=1$. The preferential positioning is observed only when the parent droplets 
are of unequal size. It should be noted here that 
the \textit{universality} of the exponent $q$ is till date questionable. For densely packed microbubbles $q=2.06 \pm 0.33$ \cite{weon2015}. For free 
bubbles $q=5.09 \pm 0.43$ and for free droplets $q=4.33\pm0.54$ \cite{je2012}. The discrepancy between these high and 
low density values has been attributed to the blocking effect due to neighboring bubbles. 

As will be discussed later, our molecular dynamics results on $q$ for phase separating fluids 
is markedly different from the above-mentioned findings. 
This is explained within the framework of centre of mass theory.

\section{Model and simulation}
For a single-component fluid a monodisperse model is considered whose particles interact via the Lennard-Jones (LJ) pair 
potential $u(r)=4\varepsilon [(\sigma/r)^{12}-(\sigma/r)^6]$, where $r$ is the scalar distance between two 
particles and $\varepsilon$ is the interaction strength. $u$ is smoothly truncated \cite{roy2012, roysoftmatter} at a 
cutoff distance $r_c$ and modified as $u_1(r)=u(r)-u(r_c)-(r-r_{_c}){{du}/{dr}}\Big|_{{r}=r_{_c}}$. 
This model exhibits a vapor--liquid transition at a critical temperature $k_B T_c \simeq 0.935 \varepsilon$ 
and a critical density $\rho_c \simeq 0.316 \sigma^3$ \cite{das2017}. 
The non-equilibrium coarsening dynamics within this model is simulated using MD in the canonical 
ensemble using the Nose-Hoover thermostat \cite{frenkel2002}. An initial configuration is prepared at a high temperature 
$5T_c$ such that it corresponds to the homogeneous phase and at time $t=0$ it is quenched inside the 
binodal to a temperature $T=0.67T_c$. The subsequent dynamics which involves the formation of droplets and 
their coalescence is monitored at a time interval of $10\tau$; $\tau=\sqrt{m\sigma^2/\varepsilon}$ is 
the LJ time unit. Periodic boundary conditions \cite{allen1987} are applied along all Cartesian directions. 
The reduced temperature $T^*$ and system size $L^*$ are defined as $T^*=k_B T/\varepsilon$, 
$L^*=L/\sigma$. $\varrho=N/V$ is the overall density of the fluid where $N$ is the total number of particles 
and $V=(L\sigma)^3$ is the volume of the cubic simulation box. Unless otherwise mentioned, data 
correspond to averaging over $6$ independent configurations.

\section{Droplet identification}
One major hurdle in quantifying the CP dynamics is to track the coalescing droplets accurately. 
This is done by using the following procedure \cite{roysoftmatter}: (i) local 
density around each particle is calculated and if this density is higher than a critical value the particle is 
marked as an element of any of the droplets. (ii) Next, depending upon the spatial distance between these 
marked particles, different droplets are identified. This method provides complete information about 
the total number of droplets in the system, the radius and volume of each droplet. The centre of mass 
and average radius of each droplet can next be calculated by assuming their spherical shape. 
From this one can easily measure $a_L/a_S$ and $R_L/R_S$ for a pair of coalescing droplets. 

\begin{figure}
\includegraphics*[width=0.45\textwidth]{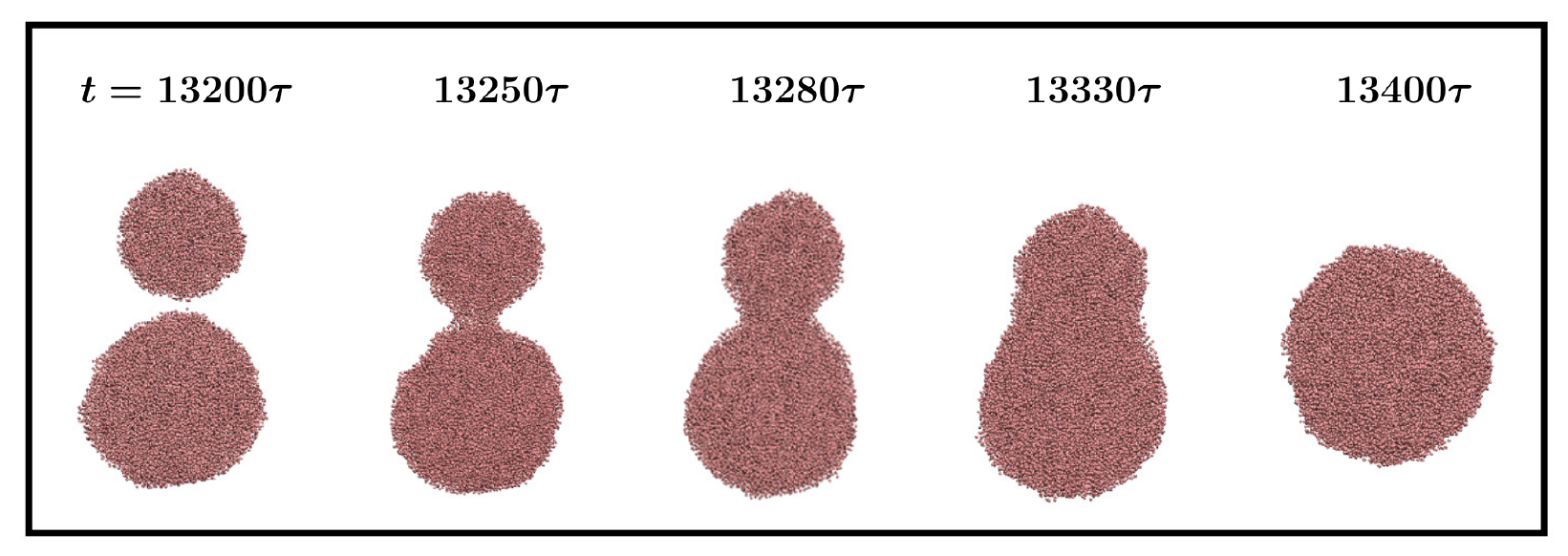}
\caption{\label{fig2}Snapshots of a binary droplet coalescence event in a phase segregating 
single-component fluid from MD simulations. 
A thin liquid bridge forms at $t=13250\tau$ which with time increases in thickness and the elongated composite 
drop relaxes to form a spherical product drop. Results correspond to $L^*=100$ and $T^*=0.67T_c^*$. }
\end{figure}

\section{Results}
In \cref{fig2}, we show an exemplary coalescence event between two droplets of different sizes produced by 
using MD simulations. As the droplets collide, a thin liquid bridge forms. With time the liquid bridge 
grows and the composite structure relaxes until at $t=13400\tau$ a single merged spherical droplet is created. 
In due course of time this merged droplet will undergo Brownian motion and will coalesce with another new droplet. 
Overall, phase separation in the whole system will progress in this way.
Note that the droplets considered in the present study obey Binder-Stauffer's \cite{binder1974, siggia1979} Brownian 
diffusion and collision mechanism, where droplets undergoing Brownian motion collide with 
each other and coalesce.

\begin{figure}
\includegraphics*[width=0.47\textwidth]{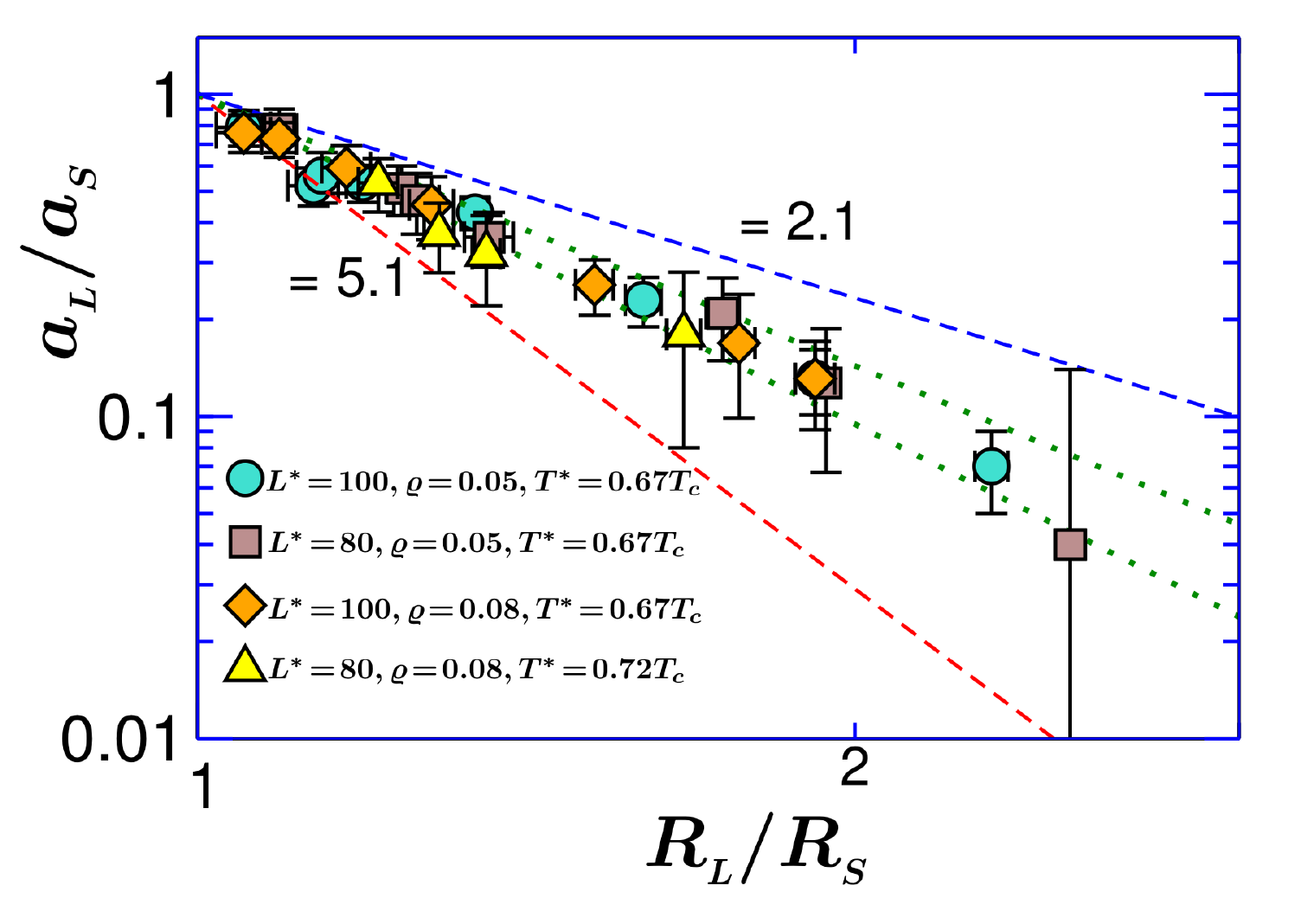}
\caption{\label{fig3}Coalescence preference: log-log plot of $a_L/a_S$ vs. the parent size 
ratio $R_L/R_S$ for droplet coalescence events in a phase segregating 
single-component fluid. Symbols correspond to simulation results for different system sizes $L^*$, 
overall densities $\rho$ of the fluid, and quench temperatures $T^*$.  The dashed lines stand for 
$q=2.1$ and $5.1$.}
\end{figure}

Next, we want to investigate the coalescence preference dynamics during such collision events. For that purpose, 
in \cref{fig3}, we plot the relative coalescence position $a_L/a_S$ 
vs. the parent size ratio $R_L/R_S$, on a double-logarithmic scale. 
Different symbols correspond to different system sizes $L^*$, overall densities $\rho$ of the fluid, 
and quench temperatures $T^*$. The chosen values of $\rho$ and $T^*$ correspond to off-critical 
quenches such that spherical droplets form \cite{royjcp2013} and the fluid phase separates via the Binder-Staufer mechanism of 
Brownian coagulation and coalescence \cite{binder1974, siggia1979}. First of all,
with increasing size inequality $R_L/R_S$, the product drop moves much closer towards the bigger parent, 
i.e., $a_L/a_S$ decreases. 
This is because increasing size inequality increases the Laplace pressure difference between the 
parents and hence the merged drop is formed more closer to the larger parent. For equal size droplets 
$(R_L/R_S=1)$ the product always forms exactly at the centre 
of the two parents, i.e., $a_L/a_S=1$. In the limiting case 
of $R_L/R_S \rightarrow \infty$ the merged drop should form over the infinite parent, i.e., 
$a_L/a_S \rightarrow 0$.
The dashed lines in \cref{fig3} 
stand for the previously reported values of $q =2.1$ and $5.1$ in the literature \cite{weon2015, je2012}. Clearly, our 
computational data in \cref{fig3} for various $L^*$, $\rho$, and $T^*$ (marked by different symbols) 
do not follow these early findings. The dotted lines mark the limiting values $2.8$ and $3.4$ 
of $q$ which the data corroborate. This yields $q \simeq 3.1 \pm 0.3$. No prominent system size 
dependence is observed in \cref{fig3}. 

Of course a broader range of abscissa, in \cref{fig3}, will lead to more 
accurate quantification. However, it should be noted that at early times during non-equilibrium phase 
separation the size dispersion of droplets is very low which results in lower values of $R_L/R_S$. 
At very late times the size dispersion increases. However, for the densities and system sizes chosen in this work 
the droplet size distribution is not broad enough to give rise to values of $R_L/R_S$ much larger than $2$.     
 
Note that from the locations of the centres of the coalescing and the merged droplets 
one can also calculate the distance $d$ between the center of the droplet appearing 
due to a coalescence event and the line of centers of two parental droplets: 
$d^2=c^2-a_L^2$; where $a_L=b(a^2+c^2-b^2)/(2ac)$ and thus study the time dependence of $d$. This 
we leave out as future exercise.

In \cref{fig3}, the obtained value of $q$ is very different from the previous reports
\cite{weon2015} $q=2.1$ and $5.1$ in the literature. We attribute this difference to the 
dominant microscopic mechanism of the coalescence event. It was already demonstrated \cite{je2012} 
that $q=5$ arises from the surface energy release during a coalescence event. Here, we propose that 
the CP dynamics in our work is governed mainly by the \textit{centre of mass theory}. Assuming that 
the product drop forms at the centre of mass (c.m) of the two coalescing parents, one obtains: 
$m_P {\vec r}_P=m_S {\vec r}_S + m_L {\vec r}_L$. Where, ${\vec r}_S$, ${\vec r}_L$, ${\vec r}_P$ are 
the position vectors of the c.m. of the smaller, larger parents and the product, respectively, 
and $m_S$, $m_L$, $m_P$ stand for their masses. 
Simple vector algebra \cite{arfkenbook} and geometric arguments lead to $(a_L/a_S) \sim (m_L/m_S)^{-1}$. 
Now, assuming that each droplet has constant spatial density within it and using the relation: 
$m=4\pi R^3/3$, one obtains the coalescence preference relation $(a_L/a_S) \sim (R_L/R_S)^{-3}$. Our MD simulation 
result $q\simeq 3.1$, in \cref{fig3}, is in excellent agreement with this scaling argument.  

\begin{figure}
\includegraphics*[width=0.46\textwidth]{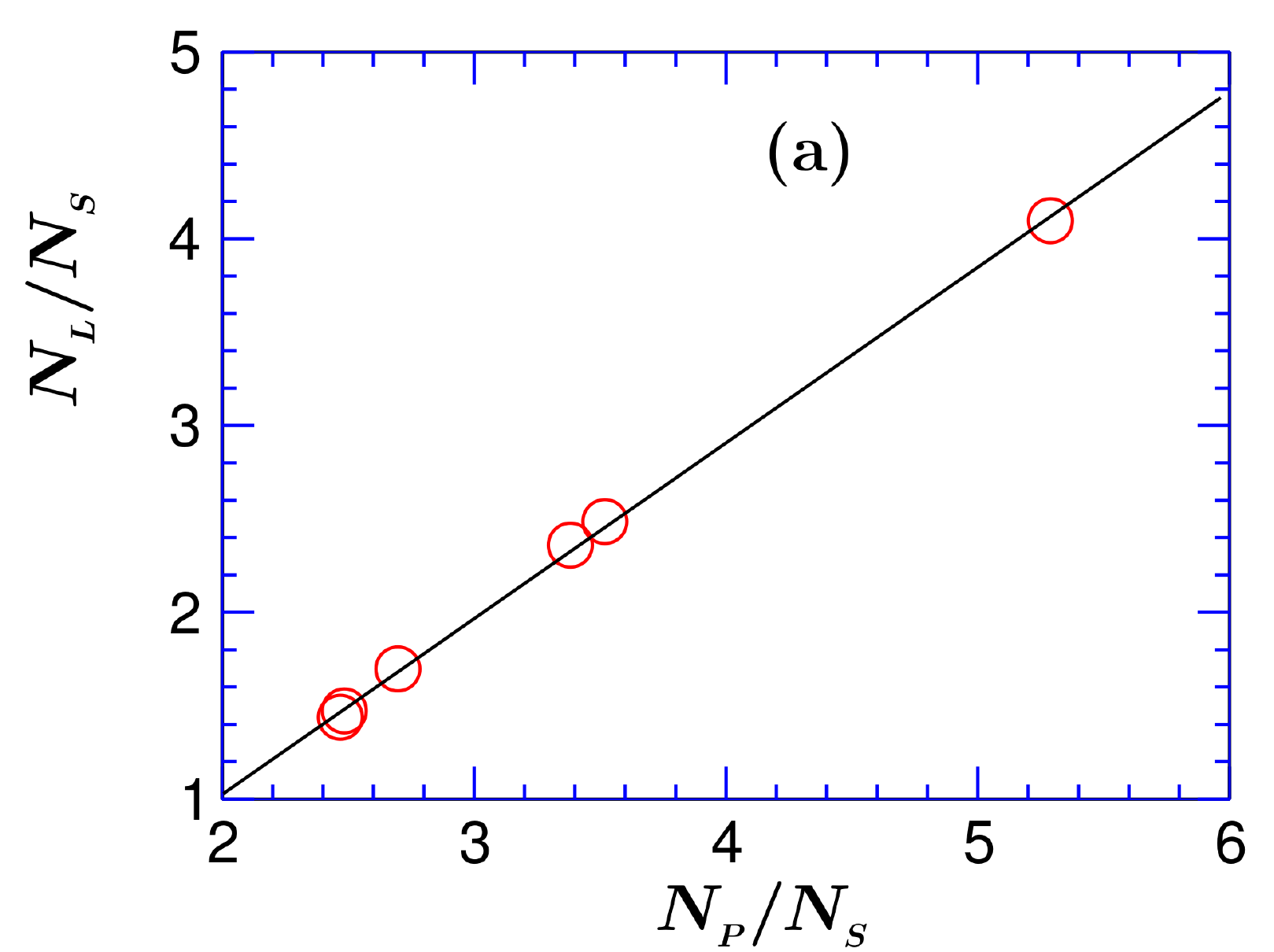}
\vskip 0.5cm
\includegraphics*[width=0.46\textwidth]{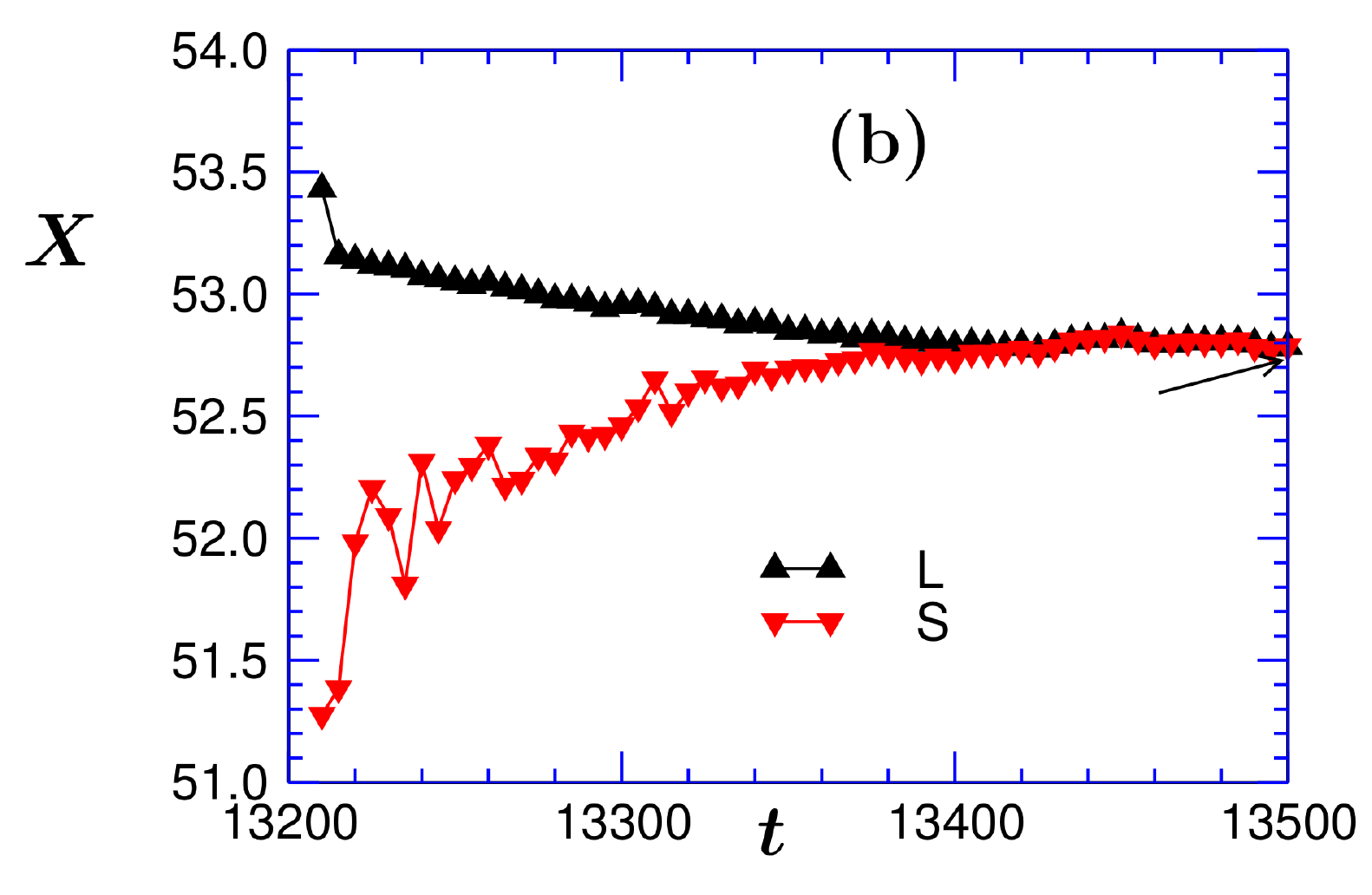}
\caption{\label{fig4}(a) Plot of the parent size ratio $N_L/N_S$ vs. the ratio $N_P/N_S$ of the sizes of the 
product and the smaller parent. Symbols refer to our simulation data and the solid line stands for 
the mass conservation law: $N_P/N_S=1+N_L/N_S$. Symbol sizes are larger than errorbars.
(b) Trajectories of two coalescing droplets. L and S correspond to the larger and the smaller ones 
respectively. With increasing time two droplet centers approach each other untill close to $13400$ 
a spherical product drop forms which does not undergo significant shape relaxation for 
$t>13400$. The product drop forms closer to the bigger parent, in accordance with the 
coalescence preference effects. Symbol sizes exceeds errorbars.}
\end{figure}

To further verify the applicability of the c.m. theory for CP dynamics in phase segregating fluids, 
we investigate various kinetic properties of the coalescence events. In \cref{fig4}(a), 
we test the law of mass conservation, $V_P=V_S+V_L$, for the coalescing 
drops considered in \cref{fig3}. Constant density approximation provides the volume of a droplet $V$ to be 
proportional to the total number of particles $N$ in the droplet. The mass conservation law thus leads to 
$N_P/N_S=1+N_L/N_S$. Data in \cref{fig4}(a) nicely corroborates with this conservation law. 
This observation is also supported by the predictions of inelastic ``\textit{sticky}'' collisions between 
coarsening droplets within the framework of Brownian coagulation theory \cite{binder1974}.

Next, in \cref{fig4}(b), we show the time evolution of the $x-$ coordinate, marked by filled symbols, 
of the centres of the two coalescing droplets considered in \cref{fig2}. The larger and smaller parents are 
marked by $L$ and $S$, respectively. Before $t\simeq13400$, centers of the individual parent droplets 
approach other during coalescence and shape relaxation of the composite object takes place. 
Finally, beyond $t\simeq 13400$, the shape of the composite structure does not change much and the spherical 
final product droplet has almost formed. As a result, after $t=13400$ locations of the two coalescing droplets 
concide with each other. It should be noted that the smaller parent (S) shifts more towards 
the bigger one (L) and the final product is formed closer to the larger parent, as expected for 
coalescence preference phenomenon. The x coordinate of the c.m. of the parent droplets 
was calculated, at time $t \simeq 13300$ when the x ordinates of the droplets just touch, 
as $x_{\text{cm}}=(x_1 R_1^3+x_2 R_2^3)/(R_1^3+R_2^3)$ and was calculated to be $\simeq 52.76$ 
(marked by arrow). 
The final product indeed forms very close to this location.  

\Cref{fig4}(a) and \cref{fig4}(b), therefore, convincingly demonstrates that the merged droplet during 
droplet coalescence, via the Brownian motion and coagulation mechanism \cite{binder1974} in a phase 
separating single-component fluid, forms at the centre-of-mass location of the parents, and it leads to 
a coalescence preference exponent $q\simeq 3$. This is in strong 
contradiction with previous studies \cite{weon2015, je2012} where the product drop does not form at the 
c.m. location. 

Droplet phase separation may involve more complex coalescence events as well. One such example being the 
coalescence induced coalescence \cite{tanaka1996}. A droplet undergoing coalescence changes its shape 
with time and during this shape relaxation process it may touch another neighboring droplet due to 
geometrical reasons. This will in turn lead to further coalescence events. This is particularly relevant when 
the average droplet density in the fluid is very high \cite{tanaka1996}. We hope that our analysis on 
rather simple binary coalescence will motivate future studies on complex droplet collisions. 
 
\section{Summary}
In summary, the coalescence preference (CP) dynamics during non-equilibrium phase separation of a 
single-component fluid has been investigated using molecular dynamics simulations and 
scaling arguments. Due to 
the off-critical order parameter of the fluid, phase separation progresses via coalescence 
of liquid droplets. To the best of our knowledge, this is the first study on CP during any 
coarsening process following a sudden temperature quench. Our results reveal that the 
location of the product drop exhibits strong dependence on the size inequality of the parents. 
With increasing size inequality the product drop forms more closer to the larger parent. 
Although this general trend is in accord with previous experiments \cite{weon2015} on 
microbubble / droplet the CP exponent, which characterizes the ``closeness'' of preferential location 
to the larger parent, is found to be strikingly different from previous reports in the literature. 
Specifically, for the present study $\simeq 3.1$ and earlier studies \cite{weon2015, je2012} reported 
$q\simeq 2.1$ and $5.1$. Such a discrepancy is attributed to the underlying mechanisms of coalescence event. 
If the motion of the merged bubble after the initial bridge formation, in a low-density system, 
is controlled by the surface energy difference between the parent and the merged bubble it leads 
to a CP exponent $q \simeq 5.1$. For a densely packed system, this exponent gets modified 
to $q\simeq 2.0$ because of the presence of multiple neighbouring bubbles which might block 
the position of the product bubble. On the other hand, in a phase separating single-component fluid droplet 
coalescence occurs via the inelastic sticky collision and coagulation and after coalescence the product droplet 
forms at the centre-of-mass (c.m.) location of the parents. Scaling arguments provide a value of $q =3$ for this 
c.m. theory with which the MD simulation data accord well. This is in contrast with droplet coalescences in 
\cite{je2012} where the merged product does not form at the c.m. location, but more closer to the 
larger parent. We hope that our study on droplet-coalescence in a phase segregating single-component 
fluid will promote future investigations on droplet coalescence in binary liquid mixures, 
coalescence of bubbles, which will in turn lead to the understanding of universality in 
coalescence preferences effects.


\begin{thebibliography}{100}
\bibitem{gaskell2007}N. Kapur and P.H. Gaskell, Phys. Rev. E. \textbf{75}, 056315 (2007). 
\bibitem{wasserfall2017}J. Wasserfall, P. Figueiredo, R. Kneer, W. Rohlfs, and P. Pischke, Phys. Rev. Fluid. \textbf{2}, 123601 (2017).
\bibitem{narhe2008}R. D. Narhe, D. A. Beysens, and Y. Pomeau, EPL \textit{81}, 4 (2008).
\bibitem{zhua2017}G. Zhua, H. Fanb, H. Huanga, and F. Duan, RSC Adv. \textbf{7}, 23954 (2017).
\bibitem{koplik2015}J. Koplik, Phys. Fluid. \textbf{27}, 082001 (2015).
\bibitem{malfreyt2013}A. Ghoufi, and P. Malfreyt, EPL \textbf{104}, 46004 (2013).
\bibitem{wilkinsin2016}M. Wilkinson, Phys. Rev. Lett. \textbf{116}, 018501 (2016). 
\bibitem{kadivar2014}E. Kadivar, EPL \textbf{106}, 24003 (2014).
\bibitem{tadrosbook}T. Tadros (editor), \textit{Emulsion Formation and Stability} 
(Wiley-VCH, Germany, 2013).
\bibitem{weon2015}Y. Kim, S.J. Lim, B. Gim and B.M. Weon, Sci. Rep. \textbf{5}, 7739 (2015).
\bibitem{je2012}B.M. Weon and J.H. Je, Phys. Rev. Lett. \textbf{108}, 224501 (2012).
\bibitem{puri2009}S. Puri and V. Wadhawan (editors), \textit{Kinetics of 
Phase Transitions} (CRC Press, Boca Raton, 2009).
\bibitem{bray2002}A.J. Bray, Adv. Phys. \textbf{51}, 481 (2002).
\bibitem{roy2012}S. Roy and S.K. Das, Rapid Comm. Phys. Rev. E \textbf{85}, 050602(R) (2012).
\bibitem{shimizu2015}R. Shimnizu and H. Tanaka, Nature Comm. \textbf{6}, 7407 (2015).
\bibitem{oprisan2017}A. Oprisas, Y. Garrabos, C. Lecoutre, and D. Baysens, Molecules \textbf{22}, 1125 (2017).
\bibitem{das2017}J. Midya and S.K. Das, Phys. Rev. Lett. \textbf{118}, 165701 (2017). 
\bibitem{binder2009}K. Binder in \textit{Kinetics of Phase Transitions} 
(CRC Press, Boca Raton, 2009), edited by S. Puri and V. Wadhawan.
\bibitem{onuki2002}A. Onuki, \textit{Phase Transition Dynamics} 
(Cambridge University Press, UK, 2002).
\bibitem{pagureva2016}N. Pagureva, S. Tcholakova, K. Rusanova, 
N. Denkova, and T. Dimitrova, Colloids and Surfaces A: Physicochemical and Engineering Aspects 
\textbf{508}, 21 (2016).
\bibitem{biance2011} A-L Biance, A. Delbos, and O. Pitois, Phys. Rev. Lett. 
\textbf{106}, 068301 (2011).
\bibitem{binder1974}K. Binder and D. Stauffer, Phys. Rev. Lett. 
\textbf{33}, 1006 (1974).
\bibitem{roysoftmatter}S. Roy and S.K. Das, Soft Matter \textbf{9}, 4178-4187 (2013).
\bibitem{das2017}J. Midya and S.K. Das, J. Chem. Phys. \textbf{146}, 044503 (2017).
\bibitem{frenkel2002}D. Frenkel, B. Smit, \textit{Understanding Molecular 
Simulations:From Algorithm to Applications} (Academic Press, San Diego, 2002). 
\bibitem{allen1987}M.P. Allen and D.J. Tildesley, \textit{Computer Simulations 
of Liquids}, Clavendon, Oxford, 1987.
\bibitem{siggia1979}E.D. Siggia, Phys. Rev. A \textbf{20}, 595 (1979).
\bibitem{royjcp2013}S. Roy and S. Das, J. Chem. Phys. \textbf{139}, 044911 (2013).
\bibitem{arfkenbook}G. Arfken, H. Weber, and F.E. Harris, \textit{Mathematical Methods for Physicists} 
(Academic Press, San Diego, 2012).
\bibitem{tanaka1996}H. Tanaka, J. Chem. Phys. \textbf{105}, 10099 (1996).


\end{thebibliography}
\end{document}